\documentstyle[12pt,fleqn,epsfig,rotate,citesort]{article}
\begin{document}

\title{\bf Intruder bands and configuration mixing
in the lead isotopes}
\vspace{2cm}

\author{\bf R.~Fossion and K.~Heyde\\
\em Department of Subatomic and Radiation Physics,\\
\em Proeftuinstraat 86, B-9000 Gent, Belgium\\
\bf G.~Thiamova\\
\em Nuclear Physics Institute, Rez, Czech Republic\\
\bf P.~Van Isacker\\
\em GANIL, BP 55027, F-14076 Caen Cedex 5, France}

\date{ }
\maketitle

\hyphenation{des-cribe heavier}

\begin{abstract}
A three-configuration mixing calculation is performed
in the context of the interacting boson model
with the aim to describe recently observed collective bands
built on low-lying $0^+$ states
in neutron-deficient lead isotopes.
The configurations that are included
correspond to the regular, spherical states
as well as two-particle two-hole
and four-particle four-hole excitations
across the $Z=82$ shell gap.

\end{abstract}

\newpage

\section{\bf Introduction}
During the last years,
ample evidence for the presence of intruder excitations,
in particular at and near closed-shell regions~\cite{hey83,woo92}
has been accumulated throughout the nuclear mass table.
Such intruder excitations,
which give rise to shape coexistence and
collective band structures,
are particularly well documented
in the $Z=50$ (Sn) and $Z=82$ (Pb) mass regions.
They can be associated
with many-particle many-hole proton excitations
across the closed shells~\cite{hey87}
but can also be studied, in an equivalent way,
using potential energy surface calculations~\cite{ben87}.

After the discovery of $0^+$ intruder excitations
in the even-even Pb nuclei,
down to mass $A=192$~\cite{dup84,dup85,dup87,den89},  
the development of recoil and recoil-decay-tagging (RDT) techniques
and of heavy-ion induced fusion-evaporation reactions~\cite{hee93}
have led to a wealth of new data
on collective bands in the Pb isotopes
that have neutron number near $N=104$.
The large number of experiments carried out over the last few years
are described in a review paper by 
Julin {\it et al.}~\cite{jul01}.
Study of the fine-structure of alpha decay
of the very neutron-deficient nuclei
has also proven to be an excellent tool~\cite{dup99}
to identify excited $0^+$ states
and was used for the first observation
of a set of three low-lying $0^+$ excitations
in $^{186}$Pb~\cite{and00,and01}.
The above two major experimental techniques
have been instrumental in exploring unexpected new phenomena
concerning intruder states and shape coexistence.

\begin{figure}
\begin{center}
\mbox{\epsfig{file=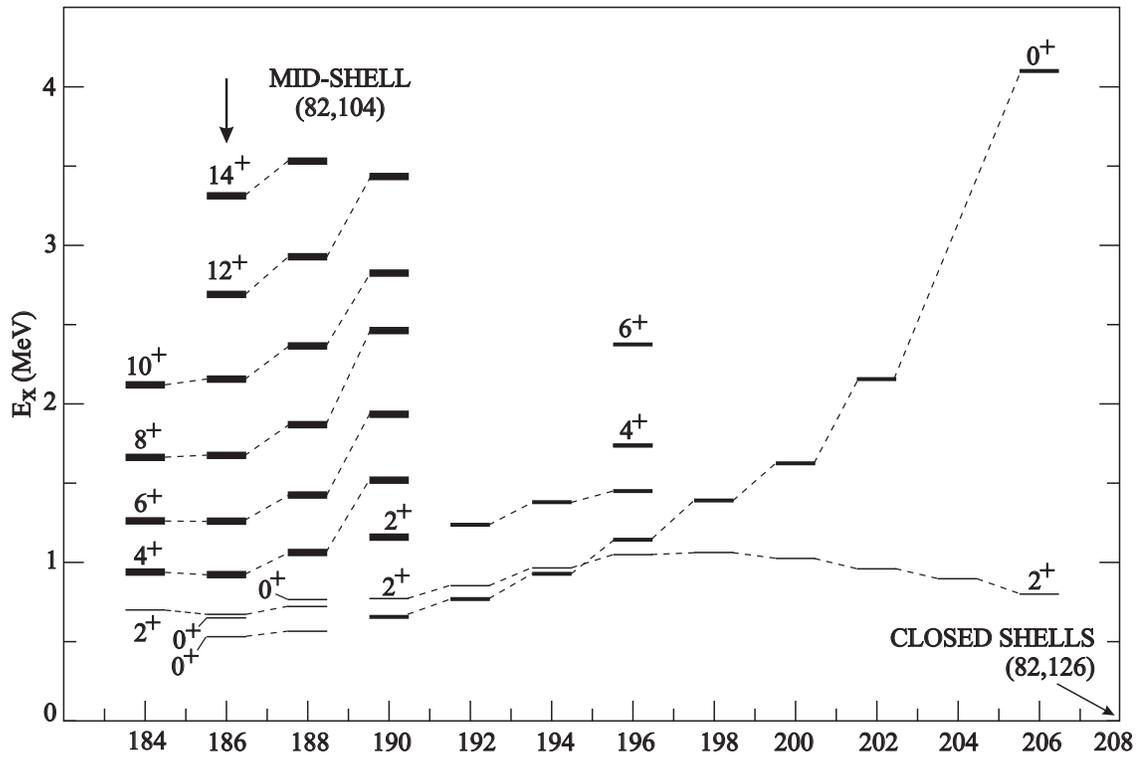,height=10.0cm}}
\end{center}
\caption{
Observed systematics of low-lying $0^+$ states
in even-even Pb nuclei.
The first-excited $2^+$ state is also given for reference
and in the mass region $184\leq A\leq190$
known band members of the yrast structure are also shown. This implies that not all observed levels are shown on this figure.
References are given in the introduction.}
%\epsfxsize=14cm
%\rotate[l]{\centerline{\mbox{\epsfbox{pbrudi.eps}}}}
\label{pbrudi}
\end{figure}

With these methods,
data have become available
on $^{188}$Pb~\cite{byr00,coz99,bij96,dra99},
$^{186}$Pb \cite{and00,and01,bax93},
$^{184}$Pb~\cite{and99,coc98}
and $^{182}$Pb~\cite{jen00}.
The experimental evidence for close-lying $0^+$ excitations
and for associated collective band structures (see Fig.~\ref{pbrudi})
has brought about the need to explore and study
the effects of mixing among these structures.
A first attempt using two-level mixing
for the low-lying $0^+$ and $2^+$ excitations in $^{190-200}$Pb
was carried out by Van Duppen {\it et al.}~\cite{dup90}.
The importance of mixing between bands
was put in a broader context by Dracoulis~\cite{dra94}
in the Os-Pt-Hg-Pb region.
More recently,
detailed two-band mixing calculations~\cite{dra98,all98}
and an attempt to perform a three-band mixing study
in $^{186}$Pb~\cite{pag02} have been reported,
albeit all starting from phenomenological band-mixing assumptions.
So, there is a clear need
to perform a detailed study of these nuclei.
We also mention that intruder $0^+$ states and associated bands
have been studied in the neutron-deficient
Pt nuclei~\cite{xu92,sew98,ced98,kin98,sor99,dra86,dav99}
down to mass $A=168$
and in the Po nuclei~\cite{ber95,you95,you97,fot97,bij95,bij98,hel96,hel99}
in an effort to distinguish
collective vibrational excitations from intruder states.

Deformed mean-field calculations~\cite{ben87,ben89,naz93,chas01,taj93}
have indicated the possibility
of obtaining oblate and prolate minima
close to the spherical ground-state configuration
in the total energy surface for the Pb nuclei
for neutron numbers close to $N=104$.
In the Hg nuclei with a ground state
corresponding to a slightly deformed oblate configuration,
a second, prolate configuration is predicted
with minimal energy near mid-shell ($N=104$),
while in the Pt nuclei
a crossing of both minima is found
and the prolate deformed minimum
becomes the lowest configuration at mid-shell.
In the Po nuclei the situation
is more complicated~\cite{oro99} with an oblate minimum
which approaches the spherical ground-state configuration
near $N=110$,
and a prolate minimum which becomes dominant in the ground
state for even lower neutron numbers.
The above studies have the drawback
that only the static potential energy surface properties
are studied.

Many-particle many-hole ($m$p-$n$h) excitations
cannot be incorporated easily
in full large-scale shell-model studies
because of the extremely large dimensions
of the model spaces involved.
These $m$p-$n$h excitations, however,
can be handled within an algebraic framework
of the Interacting Boson Model (IBM)~\cite{iac87,fra94}.
The inclusion in IBM of the simplest 2p-2h intruder excitations
in the form of two extra nucleon pairs (two bosons)
and the study of the mixing with the regular configurations
was suggested by Duval and Barrett~\cite{duv82}
and applied in the Pb region
to the Hg nuclei~\cite{bar83,bar84},
the Pt nuclei~\cite{har97}
and the Po nuclei~\cite{oro99}.
These IBM-mixing calculations, however,
generally implied the introduction of many parameters.
Calculations including up to 6p-6h excitations have been carried out, using a simplified Hamiltonian, with the aim of studying %%@
high-spin properties and backbending within the IBM configuration mixing picture by J. Jolie et al \cite{jol83,jol84}. In an effort %%@
to remove the obstacle of introducing many parameters when carrying out such IBM configuration mixing calculations with various %%@
mp-nh excitations, the symmetries connecting particle and hole bosons
have been put forward in an algebraic framework
and led to the introduction of a new label,
called {\em intruder spin} or $I$ spin~\cite{hey92}
and to applications in the Pb region~\cite{hey94}.
In this approach,
explored in detail in a series of papers~\cite{cos96,leh97,cos97,cos99},
both the particle and hole shell-model configurations
are handled as interacting particle- and hole-like $s$ and $d$ bosons.
The method also allows, in principle,
the mixing of two dynamical symmetries.

The experimental need to explore three families
each with a different intrinsic structure
(spherical, oblate and prolate
or, in a spherical shell-model language,
the 2p-2h and 4p-4h intruder excitations
next to the regular ones of 0p-0h type)
forms the starting point of the present study.
An exploratory calculation of this kind
within an algebraic framework
has already been reported~\cite{hey91,cos00}
but was confined to the $0^+$ band heads.
In this paper an IBM-mixing calculation
that describes the three different intrinsic `shape' configurations
is performed, 
and applied to the very neutron-deficient Pb nuclei.

\section{The configuration mixing model}
\subsection{Hamiltionian}
The regular (reg) 0p-0h states
are described in terms of $N$ bosons
while the intruder 2p-2h and 4p-4h states
require $N+2$ and $N+4$ bosons, respectively.
The model space thus consists
of the sum of three symmetric U(6) representations 
$[N]\oplus[N+2]\oplus[N+4]$.
The model Hamiltonian has the form
\begin{equation}
\hat H=
\hat H_{\rm reg}+
\hat H_{\rm 2p2h}+
\hat H_{\rm 4p4h}+
\hat V_{\rm mix}.
\label{hamiltonian}
\end{equation}
The different parts are 
\begin{eqnarray}
\hat H_{\rm reg}&=&
\epsilon_{\rm reg}\hat n_d+
\kappa_{\rm reg}\hat Q_{\rm reg}\cdot\hat Q_{\rm reg},
%\kappa'_{\rm reg}\hat L\cdot\hat L,
\label{reg}\\
\hat H_{\rm 2p2h}&=&
\epsilon_{\rm 2p2h}\hat n_d+
\kappa_{\rm 2p2h}\hat Q_{\rm 2p2h}\cdot\hat Q_{\rm 2p2h}+ 
%\kappa'_{\rm 2p2h}\hat L\cdot\hat L+
\Delta_{\rm 2p2h},
\label{2p-2h}\\
\hat H_{\rm 4p4h}&=&
\epsilon_{\rm 4p4h}\hat n_d+
\kappa_{\rm 4p4h}\hat Q_{\rm 4p4h}\cdot\hat Q_{\rm 4p4h}+ 
%\kappa'_{\rm 4p4h}\hat L\cdot\hat L+
\Delta_{\rm 4p4h},
\label{4p-4h}
\end{eqnarray}
with the $d$-boson number operator $\hat n_d$
and the quadrupole operator
\begin{equation}
\hat Q_i=
\left(s^\dag\tilde d+d^\dag\tilde s\right)^{(2)}+
\chi_i\left(d^\dag\tilde d\right)^{(2)},
\label{Q}
\end{equation}
where $i$ stands for a regular, 2p-2h
or 4p-4h configuration.
The parameter $\Delta_{\rm 2p2h}$
corresponds to the single-particle energy
needed to create a 2p-2h excitation,
corrected for the gain in binding energy due to pairing. 
Likewise, $\Delta_{\rm 4p4h}$ corresponds to the single-particle energy
needed to create a 4p-4h excitation (corrected for pairing),
with $\Delta_{\rm 4p4h}\approx2\Delta_{\rm 2p2h}$.
Eigensolutions depend primarily
on the ratio $\epsilon/\kappa$.
For $\epsilon\gg\kappa$ the solution is vibrational
while for $\epsilon\ll\kappa$ it is rotational.
In addition, the parameter $\chi$
in the quadrupole operator~(\ref{Q})
determines whether the solution is SU(3)
(axially deformed, $\chi=\pm\sqrt{7\over4}$)
or SO(6) (triaxially unstable, $\chi=0$).
The Hamiltonians used in~(\ref{hamiltonian})
represent the simplest parametrisation of the IBM
which encompasses its different limits,
U(5) vibrational, SU(3) rotational and SO(6) $\gamma$ soft.
Often a rotational term $\hat{L}^2$ is added
but this is not needed here.

The mixing of the different configurations
is induced by $\hat V_{\rm mix}$,
which is assumed to be of lowest possible order
in the $s$ and $d$ bosons,
\begin{equation}
\hat V_{\rm mix}=
\hat V_{\rm mix,1}+
\hat V_{\rm mix,2},
\label{mix}
\end{equation}
with
\begin{equation}
\hat V_{{\rm mix},i}=
\alpha_i\left(s^\dag s^\dag\right)+
\beta_i\left(d^\dag d^\dag\right)^{(0)}
+\mbox{hermitian conjugate},
\end{equation} 
where $i=1,2$
and $\hat V_{\rm mix,1}$
mixes the regular (0p-0h) and 2p-2h configurations
while $\hat V_{\rm mix,2}$
mixes the 2p-2h and 4p-4h configurations.
Note that care is taken
to introduce different (0p-0h)-(2p-2h)
and (2p-2h)-(4p-4h) mixing parameters,
$(\alpha_1,\beta_1)$ and $(\alpha_2,\beta_2)$, respectively,
and that there is no direct mixing
between the 0p-0h and 4p-4h configurations.
The latter statement is suggested by the observation
that the matrix element of a two-body interaction
between 0p-0h and 4p-4h states is zero.

To obtain an estimate of the relation
between $(\alpha_1,\beta_1)$ and $(\alpha_2,\beta_2)$,
assume in first approximation
that the dominant component of the 2p-2h and 4p-4h configurations
involves pairs of particles (holes)
in a single $j_{\rm p}$ ($j_{\rm h}$) shell
coupled to zero angular momentum.
For a pairing force the following ratio of mixing matrix elements
is obtained:
\begin{equation}
\frac{\left<0
\left| \hat V_{\rm pairing}\right|
j^2_{\rm p}(0)j^2_{\rm h}(0)\right>}
{\left<j^2_{\rm p}(0) j^2_{\rm h}(0)
\left|\hat V_{\rm pairing}\right|
j^4_{\rm p}(0) j^4_{\rm h}(0)\right>}=
\frac{1}{2}
\sqrt{\frac{(2j_{\rm p}+1)(2j_{\rm h}+1)}
{(2j_{\rm p}-1)(2j_{\rm h}-1)}}\approx0.5.
\label{ratio}
\end{equation}
In the limit of large shells,
the mixing between the 2p-2h and 4p-4h $0^+$ states
is thus twice as strong as
that between the ground state and the 2p-2h states.
In contrast, the corresponding ratio,
obtained with the mixing Hamiltonian~(\ref{mix})
and a U(5) vibrational classification of states,
equals
\begin{equation}
\frac{\left<s^N
\left|\hat V_{\rm mix}\right|
s^{N+2}\right>}
{\left<s^{N+2}
\left|\hat V_{\rm mix}\right|
s^{N+4}\right>}=
\frac{\alpha_1}{\alpha_2} \sqrt{\frac{(N+1)(N+2)}{(N+3)(N+4)}}.
\label{u5}
\end{equation}
In the limit of large boson number,
$\alpha_2$ should thus be twice as big as $\alpha_1$
to recover the ratio in eq.~(\ref{ratio}).
In the present application to the Pb isotopes,
the boson number $N$ is such that the square root in eq.~(\ref{u5})
equals about 0.8,
by which factor $\alpha_2$ should be reduced
to account for the finite boson number.
Another and perhaps more important fact is
that both eqs.~(\ref{ratio}) and~(\ref{u5})
are derived for a spherical configuration.
It is clear that deformation
may drastically change the ratio of mixing matrix elements
and, specifically, it may reduce both matrix elements
because of the smaller overlap
between states with different deformation.
Thus the microscopic estimate $\alpha_2\approx2\alpha_1$
cannot be adhered to rigidly
and several calculations
corresponding to different combinations of mixing matrix elements
must be explored.

\subsection{Parametrisation from known experimental data}
\subsubsection{The regular configuration}
In the even-even Pb nuclei,
$126-N$ valence neutrons form a ground state
which is predominantly of seniority zero, $v=0$.
This can be viewed as a condensate
of neutron pairs coupled to $0^+$.
Excited states are first built from one broken pair
($v=2$) with $2^+,4^+,6^+,\ldots$,
next two broken pairs ($v=4$) $0^+,2^+,3^+,4^+,\ldots$
and so on.
To describe these spherical states in an IBM context,
we consider the Hamiltonian~(\ref{reg})
with a spherical structure [symmetry limit U(5)]
and which has $\epsilon_{\rm reg}$ as the only parameter.
We fix this parameter to the $2^+_1$ state
of the heavier Pb isotopes $A=198$ to 204,
where the $m$p-$n$h intruder states are high in energy
and have little effect on the low-excited spherical states (see fig.~\ref{pbrudi}).
We do not consider the one broken pair $4^+$ ($g$ boson)
or $6^+$ ($i$ boson),\ldots states,
which can only be described within extended boson models.
A constant Hamiltonian for this regular configuration,
$\hat H_{\rm reg}=\epsilon_{\rm reg}\hat n_d$,
yields a constant excitation energy for the $2^+_1$ state,
independent of the number of bosons.
The excitation energy of the observed states, however,
changes slightly with mass number $A$
(because of shell effects).
We choose $\epsilon_{\rm reg}=0.9$ MeV;
this value is a compromise
between a $2^+_1$ state which is somewhat too low
and other states
that increase too rapidly in energy with angular momentum.
A better fixing of $\epsilon_{\rm reg}$ could have been obtained
by a variation of $\epsilon_{\rm reg}$
linear with the number of bosons.
For the sake of simplicity this is not done here.

\subsubsection{The intruder excitations: 2p-2h and 4p-4h configurations}
The spectroscopic information
on the neutron-deficient Pb isotopes is limited.
We use $I$-spin symmetry arguments
in order to find the IBM parameters
of the two intruder configurations in Pb
from corresponding bands in neighbouring nuclei
where more experimental information is available. 

\begin{figure}
\begin{center}
\mbox{\epsfig{file=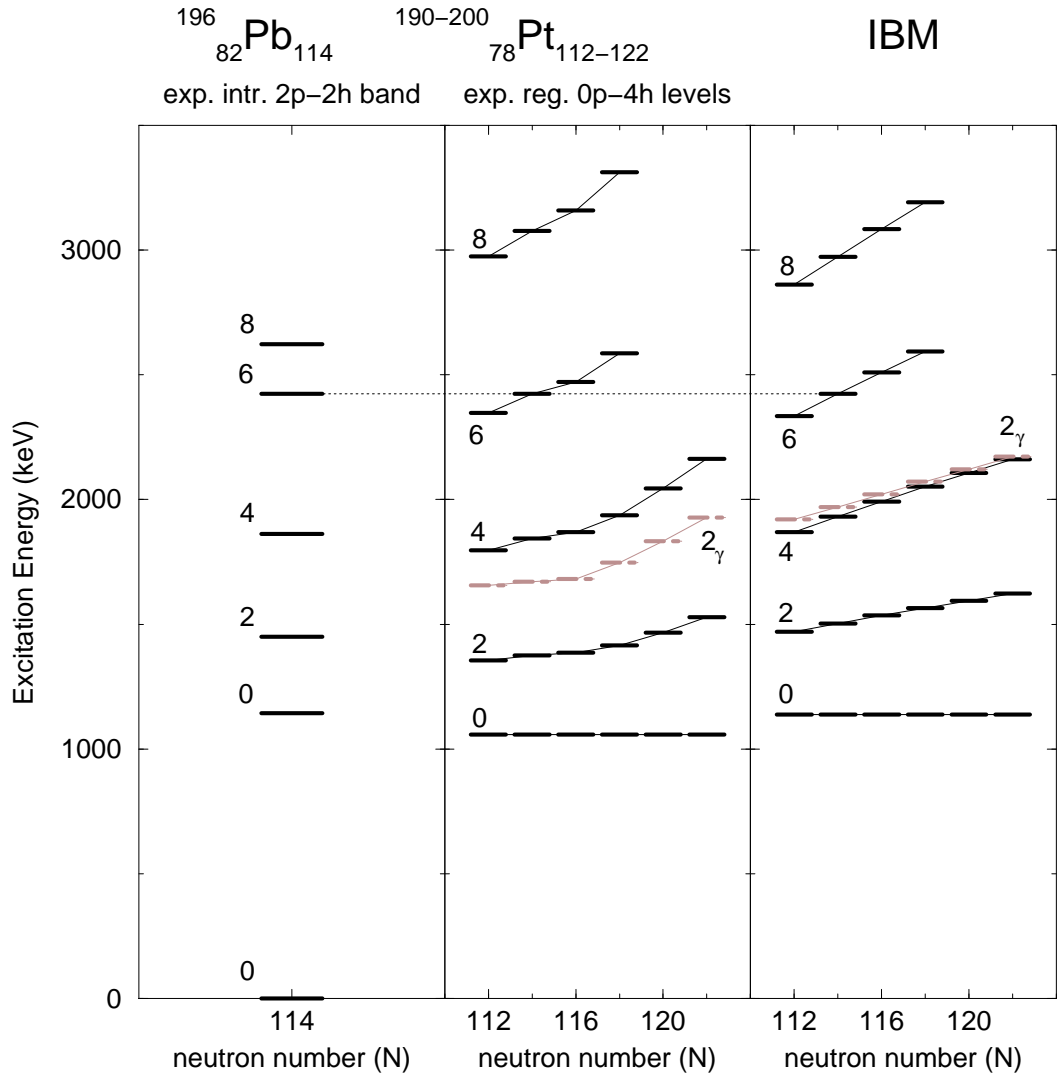,height=14.0cm,angle=-90}}
\end{center}
\caption{
Left panel: Experimental~\protect\cite{pen87} excitation energies
of the 2p-2h intruder band in $^{196}$Pb
relative to the regular $0_1^+$ ground state.
Middle panel: Experimental
\protect\cite{xu92,sew98,ced98,kin98,sor99,har97,dra91,dup99,ced90,dav99,voi90,pop97,dra86}
regular (0p-4h) ground-state bands
in the Pt isotopes with $A=190$ to 200.
Right panel: IBM calculation for the Pt isotopes in the same mass range
with parameters $\epsilon=0.51$ MeV, $\kappa=-0.0140$ MeV and $\chi=0.515$ (kept constant for all isotopes).
The Pb and Pt bands are suggested
as members of an $I=1$ intruder-analog multiplet.
Energies in Pt and IBM are normalized
to the $6^+$ excitation energy in the $N=114$ isotope
and by requiring a constant $0^+$ energy in other isotopes.}
\label{pb_2p2h}
\end{figure}

Before deriving the parameters
for the 2p-2h configuration in Pb,
we first show in fig.~\ref{pb_2p2h}
the correspondence between
the experimental 2p-2h intruder band
in $^{196}$Pb---the
only Pb isotope where it is known---(left panel)
and the intruder analogue regular 0p-4h ground-state band
in $^{192}$Pt (middle panel).
This 0p-4h band is also shown for other Pt isotopes
from mass numbers $A=190$ to 200,
where the coexisting 2p-6h intruder configuration
lies high in energy
and has little influence on the 0p-4h band.
The right-hand panel of fig.~\ref{pb_2p2h}
shows an IBM calculation of the 0p-4h band in the Pt isotopes
which, using a constant Hamiltonian
\begin{equation}
\hat H =
\epsilon\hat n_d+\kappa\hat Q\cdot\hat Q,
\label{H}
\end{equation}
with $\hat Q$ given in eq.~(\ref{Q}).
The band head of the $\gamma$ band, the $2^+_2$ level,
is also taken into account in order to restrict the possible parameter values.
The parameter $\chi$ is fixed in order
to reproduce the observed ratio $R$
\begin{equation}
R=\frac{B(E2:2^+_2\rightarrow0^+_1)}
{B(E2:2^+_1\rightarrow0^+_1)},
\label{R}
\end{equation}
which is 0.00732 in $^{192}$Pt~\cite{nds98}
and 0.00485 in $^{194}$Pt~\cite{nds96}.
Theoretical values are 0.00711 and 0.00500, respectively.
It is seen that the experimental spectra for Pb and Pt
are very similar up to $6^+$.
Intruder spin symmetry appears to be a good approximation,
and so it seems reasonable to use
the 0p-4h parameters of Pt
to describe the Pb intruder 2p-2h configuration.
 
\begin{figure}
\begin{center}
\mbox{\epsfig{file=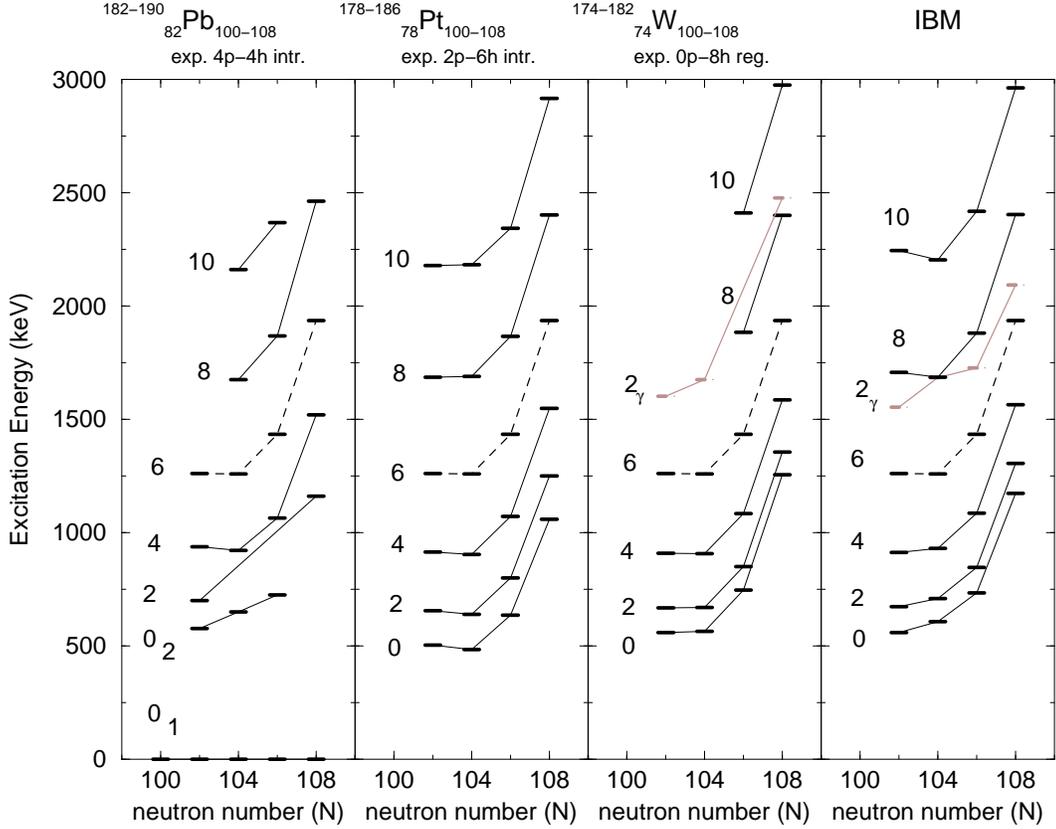,height=14.0cm,angle=-90}}
\end{center}
\caption{
Left panel: Experimental
\protect\cite{dup84,dup85,dup87,den89,hee93,jul01,dup99,and00,and01,byr00,coz99,bij96,bax93,and99,coc98,jen0%%@
0,dup90,dra98,dra99,all98,pen87,bal94,kan86,dra94,pag02}
4p-4h intruder bands
in the Pb isotopes with $A=182$ to 190.
Second panel: Experimental 2p-6h intruder bands
in the Pt isotopes with $A=178$ to 186.
Third panel: Experimental~\protect\cite{kib01,wu89}
regular (0p-8h) ground-state bands
in the W isotopes with $A=174$ to 182.
Right panel: IBM calculation for the W isotopes in the same mass range
with parameters $\epsilon=0.55$ MeV, $\kappa=-0.020$ MeV and $\chi=-0.68$ (kept constant for all isotopes).
The Pb, Pt and W bands are suggested
as members of an $I=2$ intruder-analog multiplet.
Energies in Pt, W and IBM are normalized
to the $6^+$ excitation energy in the corresponding Pb isotope.}
\label{pb_pt_w6}
\end{figure}

In the same way we derive the parameters
for the 4p-4h configuration in Pb.
In fig.~\ref{pb_pt_w6}
we show the observed analogue bands of the 4p-4h intruder configuration
in the Pb isotopes (left panel),
the 2p-6h intruder configuration
in the Pt isotopes (second panel)
and the regular 0p-8h ground-state configuration
in the W isotopes (third panel). 
In the right-hand panel of fig.~\ref{pb_pt_w6},
an IBM calculation for this 0p-8h band in the W isotopes is shown,
with the Hamiltonian~(\ref{H})
again with constant parameters for all isotopes.
This calculation also includes the $2^+_2$ level,
the bandhead of the $\gamma$ band.
The parameter $\chi$ is fixed
from the experimental ratio~(\ref{R}) in $^{182}$W,
which lies in the range 0.0252--0.0295
according to the different references in~\cite{nds95};
the calculated ratio is 0.0276.
Again, $I$-spin symmetry seems to be valid in this case
and we use the 0p-8h parameters of W
to describe the 4p-4h configuration in Pb.

\subsection{Three-configuration mixing in the Pb isotopes}
\label{3config}
With the three parameter sets ($\varepsilon_i,\kappa_i,\chi_i$) determined previously 
for the regular, 2p-2h and 4p-4h configurations,
a mixing must now be supplied
that reproduces the experimental features
of the three-configuration coexistence in the Pb isotopes (the extra parameters $\alpha_i,\beta_i,\Delta_i$).
The final set of parameters is given in table~\ref{table}. 
In table~\ref{table2}, we present results on the mixing of the three configurations (reg, 2p-2h and 4p-4h, respectively) in the first three $0^+$ excited states. One clearly notices a rapid stabilization, close to 100\%, for the $0^+_1$  state at the spherical configuration. The $0^+_2$ and $0^+_3$ states indicate a strong mixing between the 2p-2h and 4p-4h configurations. Rather quickly, a dominance of the 2p-2h configuration is observed from $A=186$ up to heavier masses in the $0^+_2$ state. In table~\ref{table3}, we identify, for $^{186}_{82}Pb_{104}$, the bands which contain mainly the 2p-2h (called "2p-2h" band) and the 4p-4h (called "4p-4h" band) configuration for a particular $J^\pi_i$ value. One notices the "4p-4h" band is related to the $0^+_2$  band head and follows up the $2_1^+, 4^+_1, 6^+_1, 8^+_1$ states, whereas the "2p-2h" band is connected to the $0^+_3$ band head and follows up the $2^+_2, 4^+_2, 6^+_2, 8^+_2$ states, respectively. The results in tables~\ref{table2} and \ref{table3} are particularly illuminating in showing the mixing effects within the IBM configuration mixing.

Note that with constant parameters for all Pb isotopes
identical results are obtained for corresponding isotopes 
that are symmetric around the middle of the shell.

\begin{figure}
\begin{center}
\mbox{\epsfig{file=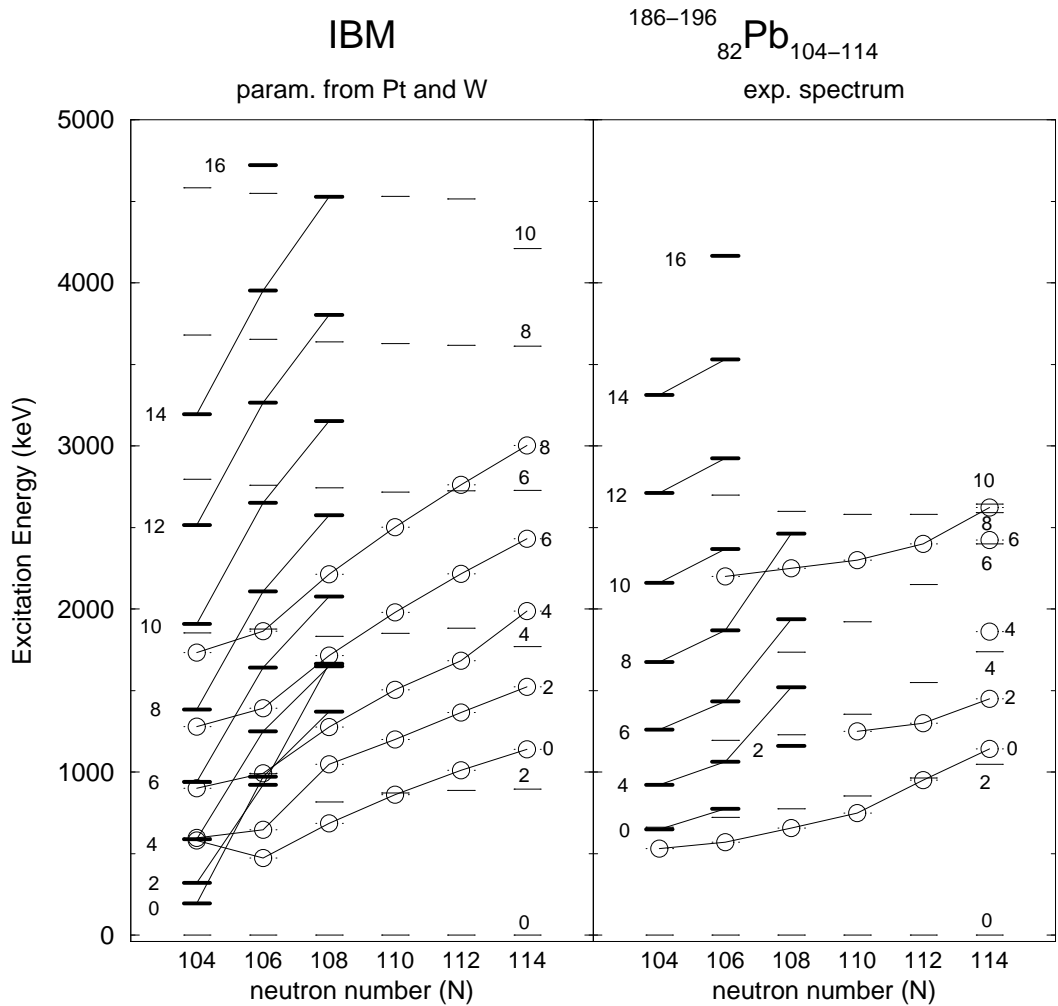,height=14.0cm,angle=-90}}
\end{center}
\caption{
Right panel: Experimental
\protect\cite{dup84,dup85,dup87,den89,hee93,jul01,dup99,and00,and01,byr00,coz99,bij96,bax93,and99,coc98,jen0%%@
0,dup90,dra98,dra99,all98,pen87,bal94,kan86,dra94,pag02}
spectra of the Pb isotopes with $A=186$ to 196
containing a regular ground-state spherical band (thin horizontal lines),
a 2p-2h intruder band (open circles)
and a 4p-4h intruder band (thick horizontal lines).
Left panel: IBM calculation for the Pb isotopes in the same mass region,
with parameters determined from results obtained in figs.~\ref{pbrudi}-\ref{pb_pt_w6}, as given in table~\protect\ref{table}.}
\label{pb_sph}
\end{figure}

The result of such a mixing calculation
is shown in fig.~\ref{pb_sph} (left panel).
From the comparison with the experimental situation,
see fig.~\ref{pb_sph} (right panel), the slope with which
the intruder states gain binding energy towards midshell
is too steep,
an effect that is independent of the mixing.
Apparently, $I$-spin symmetry is not satisfied completely.
A possible cause for these differences
between the experimental and theoretical band structures
is the fact that the IBM parameters are obtained from separate calculations
for the 2p-2h intruder excitations (fig.~\ref{pb_2p2h})
and 4p-4h intruder excitations (fig.~\ref{pb_pt_w6}).
In the final calculation for the even-even Pb nuclei,
which includes the 2p-2h and 4p-4h intruder families
together with the regular spherical configuration,
the extra mixing effects will cause specific perturbations
that may show up as a breaking of $I$-spin symmetry
and may also affect the $A$-dependence of the intruder band energies. 

Possible improvements can be expected from a more elaborate mixing study,
treating {\em all} known data
for the neutron-deficient Pb isotopes
and re\-levant nuclei in the $Z=82,N=126$ region.
A further constraint would also be imposed by the condition
that the lowest-lying proton 2p-2h $0^+$ excitations
approach the observed energy at $E_{\rm x}\approx4$ MeV in $^{208}$Pb.
Such extensive calculations,
aiming to describe besides energy spectra and band structures
also known electromagnetic (mostly E2 and E0) properties,
isotopic and isomeric shifts and nuclear ground-state binding energies,
will be pursued in the near future. 

Very recently, calculations have been carried out by C. Vargas et al. \cite{var02}, in order to determine total energy surfaces, starting from the IBM parameters derived in the present paper (see also table~\ref{table}), using boson coherent states.

\section{Conclusion}
We have explored in the present paper
the possibility to describe the co\-existence and mixing
of three different families of excitations,
corresponding to different intrinsic structures
as observed in the Pb isotopes.
In a mean-field approach
the three configurations correspond
to spherical, oblate and prolate shapes
while in a shell-model description they arise
as regular 0p-0h states
and 2p-2h and 4p-4h excitations across the $Z=82$ proton shell.
Previous studies of such three-configuration systems
were limited to the $0^+$ band heads (without the associated bands)
either in a mean-field~\cite{ben87,ben89,naz93,chas01,oro99}
or an IBM~\cite{hey91,cos00} approach,
or, in $^{186}$Pb,
to a phenomenological band-mixing calculation~\cite{pag02}.

The present description is based
on the IBM-configuration mixing approach
which was originally proposed
for two configurations~\cite{duv82,bar83,bar84,har97}.
In an IBM framework the three configurations are described
as $N$-, $(N+2)$- and $(N+4)$-boson states.
To reduce the number of parameters
that appear in such a configuration mixing calculation,
use is made of the concept of intruder-spin symmetry~\cite{hey92},
relating configurations
with different numbers of particle ($N_{\rm p}$)
and hole ($N_{\rm h}$) bosons (i.e., fermion pairs),
but with a constant sum $N_{\rm p}+N_{\rm h}=N.$
In this way experimental excitation energies in adjacent Pt and W nuclei
are used to fix the essential IBM parameters.
The final result is a reasonably good description of the observed bands
and their variation with mass number $A$.
As outlined at the end of section~\ref{3config},
ways to improve the present results
will need extensive calculations
spanning the region of neutron deficient W, Os, Pt, Hg, Pb, Po and Rn nuclei
as well as nuclei near the $Z=82,N=126$ doubly-closed shell.

The authors are grateful
to M.~Huyse, W.~Nazarewicz and P.~Van Duppen
for stimulating discussions on the shell-model interpretation
and to J.~Jolie and J.L.~Wood for help
in building the various parts of this description.
Financial support of the IWT,
the ``FWO-Vlaanderen''
and the DWTC (grant IUAP \#P5/07) is acknowledged. RF also receives financial support from a Marie Curie Fellowship of the European Community (contract number 2000-00084).

\newpage
\pagestyle{empty}

\begin{table}
\caption{Parameters of the IBM-calculation
of the three coexisting configurations in the Pb isotopes, as shown in fig.~\ref{pb_sph}. 
All parameters are in MeV, except for $\chi$,
which is dimensionless.}
\begin{center}
\begin{tabular}{r|ccccc}
\hline
configuration& $\epsilon$& $\kappa$& $\chi$& $\Delta$& mixing\\  
\hline
regular&    $0.9$&         $0$&      ---&      ---&       \\
&&&&&                             $\alpha_1=\beta_1=0.015$\\    
2p-2h  &   $0.51$&    $-0.014$&  $0.515$&   $1.88$&       \\ 
&&&&&                             $\alpha_2=\beta_2=0.030$\\
4p-4h  &   $0.55$&    $-0.020$& $-0.680$&   $4.00$&       \\
\hline
\end{tabular}
\end{center}
\label{table}
\end{table}

\newpage

\pagebreak

\begin{table}
\caption{Magnitudes (in percentages, \%) of the three different configurations (reg, 2p-2h, 4p-4h) in the first three $0_i^+$ states of the Pb isotopes $186 \leq A \leq 196$, in the IBM calculation of fig.~\ref{pb_sph} (left panel). }
\begin{center}
\begin{tabular}{c|c|c|c|c|c|c|c}

  \hline
  $0_i^+$ & configurations & $^{186}_{82}Pb_{104}$ & $^{188}_{82}Pb_{106}$ & $^{190}_{82}$Pb$_{108}$ & $^{192}_{82}$Pb$_{110}$ & $^{194}_{82}$Pb$_{112}$ & $^{196}_{82}$Pb$_{114}$ \\
  \hline
         & reg & 75 & 93 & 96 & 98 & 99 & 99 \\
  $0_1^+$ & 2p-2h & 18 & 7 & 4 & 2 & 1 & 1 \\
         & 4p-4h & 7 & 1 & 0 & 0 & 0 & 0 \\ \hline
         & reg & 20 & 6 & 4 & 2 & 2 & 1 \\
  $0_2^+$ & 2p-2h & 27 & 67 & 82 & 89 & 92 & 95 \\
         & 4p-4h & 52 & 27 & 14 & 9 & 6 & 4 \\ \hline
         & reg & 4 & 1 & 1 & - & - & - \\
  $0_3^+$ & 2p-2h & 51 & 28 & 43 & - & - & - \\
         & 4p-4h & 45 & 71 & 57 & - & - & - \\
  \hline
\end{tabular}
\end{center}\label{table2}
\end{table}

\newpage

\begin{table}
\caption{The magnitudes (given in percentages, \%) of the three different configurations (reg, 2p-2h, 4p-4h) in the "2p-2h" and "4p-4h" assigned bands of $^{186}_{82}Pb_{104},$ as shown in the IBM calculation of fig.~\ref{pb_sph} (left panel)}
\center
\begin{tabular}{c|c|c|c||c|c|c|c}
  \hline
\multicolumn{4}{c||} {\mbox {\bf  "2p-2h band"}} & \multicolumn{4}{c} {\mbox {\bf "4p-4h band"}} \\
  \hline
     & reg & 2p-2h & 4p-4h &   & reg & 2p-2h & 4p-4h \\
\hline
  $0_3^+$ & 4 & 51 & 45 & $0_2^+$ & 20 & 27 & 52 \\
  $2_2^+$ & 5 & 65 & 30 & $2_1^+$ & 1 & 32 & 67 \\
  $4_2^+$ & 1 & 77 & 22 & $4_1^+$ & 0 & 22 & 78 \\
  $6_2^+$ & 1 & 81 & 18 & $6_1^+$ & 0 & 17 & 83 \\
  $8_2^+$ & 0 & 83 & 16 & $8_1^+$ & 0 & 16 & 84 \\
  
  \hline
\end{tabular}\label{table3}
\end{table}

%\newpage

\end{document}